# 1D goes 2D: A Kosterlitz Thouless transition in superconducting arrays of 4-Angstrom carbon nanotubes

Zhe Wang[1], Wu Shi[1], Hang Xie[1], Ting Zhang[1], Ning Wang[1], Zikang Tang[1], Xixiang Zhang[1,2], Rolf Lortz[1], and Ping Sheng[1*]

**We report superconducting resistive transition characteristics for array(s) of coupled 4-Angstrom single wall carbon nanotubes embedded in aluminophosphate-five (AFI) zeolite. The transition was observed to initiate at 15K with a slow resistance decrease switching to a sharp, order of magnitude drop between 7.5-6.0K. The transition has strong (anisotropic) magnetic field dependence. Differential resistance versus current (voltage) measurements indicate that the establishment of coherence proceeds in stages as the temperature is lowered below 15K. In particular, the sharp resistance drop and its attendant nonlinear IV characteristics are consistent with the manifestations of a Kosterlitz-Thouless (KT) transition that establishes quasi long range order in the plane transverse to the c-axis of the nanotubes, leading to an inhomogeneous system comprising 3D superconducting regions connected by weak links. Global coherence is established at below 5K with the appearance of a well-defined supercurrent gap at 2K.**

Superconductivity in carbon nanotubes has been a controversial issue because carbon is not known to be a superconducting element, and if there is indeed superconducting tendency in carbon nanotubes[1], its manifestation could be quenched by long wavelength thermal fluctuations as well as by the Peierls distortion that favors an insulating ground state. In this context the earlier report on the Meissner effect in 4-Angstrom carbon nanotube-zeolite composites[2] and the more recent observation of their superconducting specific heat signals[3] have only deepened the mystery on the specific manner in which the nanotube superconductivity comes into being, and on whether there can be a sharp superconducting resistive transition that is usually taken to be the hallmark of a superconductor. In this work we show that by devising strategy to make surface electrical contacts to the samples that are separated by only 100nm, reliable and repeatable observations of the superconducting resistive transition can be obtained. The overall physical picture that emerges is that of a coupled Josephson array consisting of aligned nanotubes crossing over from an individually fluctuating 1D system to a coherent 3D superconductor, mediated by a Kosterlitz-Thouless (KT) transition[4] which establishes quasi long range order in the lateral plane perpendicular to the c-axis of the nanotubes. The attainment of overall coherence across the measuring electrodes (denoted as global coherence in this work) is seen at 5K and below, accompanied by the appearance of a well-defined (differential resistance) supercurrent gap at 2K.

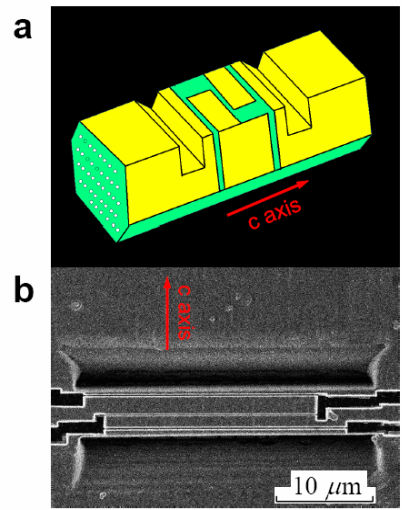

**Figure 1 | Geometry of the electrical contacts.**
**a,** Cartoon picture of AFI zeolite crystal with the 4-probe contact geometry. Yellow denotes gold and green denotes AFI crystal surface exposed by FIB etching. Nanotubes, denoted schematically by open circles, are aligned along the c axis.
**b,** A SEM image of the sample with the contact geometry illustrated in **a**. The dark areas are the deep troughs. The 100nm-width line etched by the FIB is visible in the center of the image. It separates the two surface electrodes.

[1]Department of Physics and William Mong Institute of Nano Science and Technology, HKUST, Clear Water Bay, Kowloon, Hong Kong, China.
[2]Present address: Research and Development, King Abdullah University of Science and Technology, Thuwal, Saudi Arabia.
*Corresponding author. Email address: sheng@ust.hk.

**Electrical contact geometry**
Figure 1 shows both a cartoon picture (1a) of the AFI zeolite crystal with the 4-probe contact geometry, as well as a scanning electron microscope (SEM) image of an actual sample (1b). Here the crystal was prepared by first cutting two troughs in an AFI crystal (50×50×500μm) with focused ion beam (FIB, Seiko SMI2050). The troughs are separated by a 5μm slice that is perpendicular to the c-axis (Fig. 1a).

The straight pores of the AFI zeolite are aligned along the c-axis with a center-to-center separation of 1.37nm. They form a 2D close-packed triangular lattice in the plane transverse to the c-axis. The inner diameter of the pores, after discounting the size of the oxygen atoms lining the walls, is 0.7nm[5]. The 4-Angstrom nanotubes are embedded in the AFI zeolite pores. There can be three types of 4-Angstrom carbon nanotubes that are consistent with optical and Raman data[6-10]: (5,0), (4,2) and (3,3). We attribute the superconducting behavior to the (5,0) nanotubes.

The AFI zeolite crystal with embedded nanotubes was sputtered with 50 nm of Ti and 150nm of Au. The electrical contact geometry was subsequently delineated by using the FIB to remove the Au/Ti film in a pre-designed pattern, shown in Fig. 1b. Here the outer electrodes make end contacts to the nanotubes, whereas the two inner electrodes, each about 2μm wide, are separated by 100nm and are on the surface of the AFI crystal. As the nanotubes are only ~1nm below the surface, which is imperfect in any case, the surface contact electrodes enables the measurement of electrical characteristics transverse to the c-axis of the nanotubes, i.e., the contact resistance in this case is the transverse resistance. We have carried out measurements using both the four-probe (with the outer contacts as the current electrodes) and the two-probe geometry. In the latter the two surface-contact electrodes were used. In what follows we show results mostly done in the two-probe geometry because of the importance of the transverse resistance in the KT transition mechanism. Use of four-probe geometry is noted whenever such result is presented,

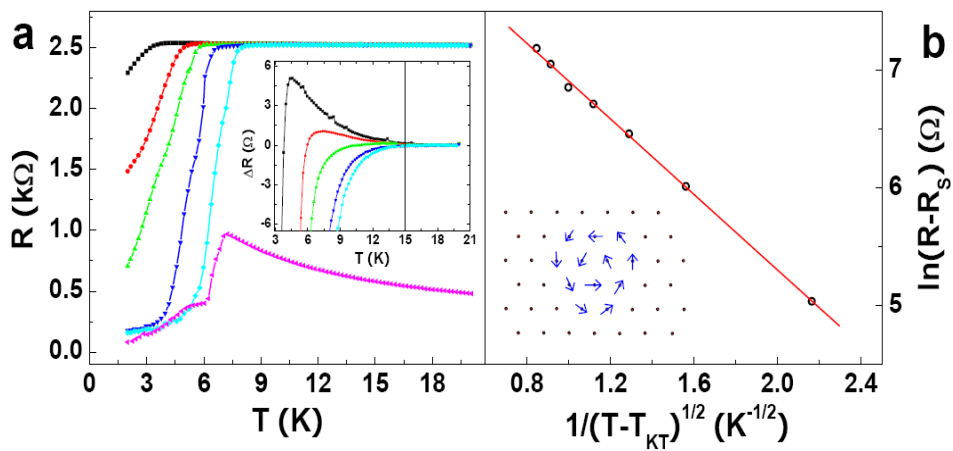

**Figure 2 | Temperature dependence of resistance measured with a magnetic field applied perpendicular to the c axis.**
**a,** Resistance measured at 1μA plotted as a function of temperature. At zero field, the sharp drop initiates at ~7.5K. The transition is pushed to lower temperatures with applied magnetic field, but still not completely quenches at 9T. This sharp drop in resistance is attributed to a KT transition in the plane perpendicular to the c axis. The magenta curve is the four leads data at zero field, with the other curves measured with the two probe geometry. From right to left, the light blue curve denotes 0T, dark blue for 1T, green for 3T, red for 5T, and black for 9T. The same color coding applies to the curves in the inset. The nonmetallic temperature dependence of the four-probe result at $T>7.5K$ is indicative of the discontinuities of the nanotubes, leading to hopping conduction and weak links. **Inset:** A magnified view of the upper section of the resistance vs. temperature curves from 3K to 20K with the straight-line asymptote above 17K subtracted. The superconducting transition clearly begins at 15K.
**b,** Under zero field, the $R(T)$ associated with a KT transition is predicted to behave as $\ln(R-R_s) \propto (T-T_{KT})^{-1/2}$ for $T > T_{KT}$, denoted by the solid red line. Very good agreement is seen with the data, denoted by open circles. Here $T_{KT}=6.17K$ and $R_s=1.06k\Omega$ is the series resistance associated with the lower plateau value in Fig. 4. It arises from the weak links connecting the different superconducting regions at low temperatures. **Inset:** A schematic picture of the lateral plane perpendicular to the c-axis, in which the KT transition occurs. Here each dot represents an end-view of a segment of the 1D element with a constant phase. The vortex is indicated by closed rings of arrows, representing unit vectors whose directions are given by the phase (angles) of the 1D element.

mainly to delineate the electrical anisotropy through comparison with the two-probe results.

## Temperature and magnetic field dependences of resistance

In Fig. 2 we plot the resistance measured as a function of temperature. The most prominent feature is the sharp drop starting at ~7.5K, which moves to lower temperatures with applied magnetic field (perpendicular to the c-axis in this case). The four-probe result, measured at zero field, is also displayed for comparison. While the sharp drop feature is preserved in the four-probe data, there is also a kink at 6K, with a small plateau extending about half a degree below that. In the inset to Fig. 2a we show an enlarged upper section of the curves from 3-20K. For $T > 17K$ the curves are very good straight lines with a slight negative slope; the data shown in the inset are measured relative to this straight-line asymptote, extended to lower temperatures.

There are several notable features of the data. First, the fact that the transition is sensitive to the magnetic field, in the range of 1T or above, means that the superconducting behavior must originate from a fairly extensive array of coupled nanotubes at least a few tens of nanometers in its lateral size[11]. Similar behavior as displayed in Fig. 2a has been observed in three different samples, with the same drop at 7.5K, although the absolute values of the measured resistances are different. Second, the inset clearly shows the initiation of the whole transition process starts at 15K. While the resistance changes are small, their magnetic field dependence is unmistakable. Third, the 4-probe data shows that there could be two stages in reaching global coherence. This point is reinforced by the IV characteristics as shown below. In Fig. 2b we show the zero field resistance versus temperature data to be in excellent agreement with the theoretical prediction based on our interpretation of the 7.5 K transition as a KT transition with a . This is a key point that will be further elaborated, together with the nonlinear IV characteristics below 7.5K that are consistent with the manifestations of the KT transition.

In Fig. 3a we show the magnetoresistance (MR) of the same sample at different temperatures. Here the magnetic field is applied perpendicular to the c-axis. At 2K, there is a kink at ~2T. For comparison, four-probe data are also shown. It is instructive that the measured resistances for the two sets of data are almost identical below 2T, but diverge above that, i.e., the contact resistance of the surface electrodes can be turned on and off through either temperature or magnetic field. It implies that at $T = 2K$ and field below 2T the transverse (contact) resistance is markedly decreased, apparently due to increased 3D coherence[12]. Figure 3b shows the observed magnetoresistance anisotropy in a separate but similar sample measured at 5T, the maximum in terms of the observed anisotropy. The anisotropy means that a magnetic field perpendicular to the c-axis is more effective in suppressing the superconducting behavior than the same field applied parallel to the c-axis. This is reasonable since the magnetic susceptibility should be larger for the perpendicular field, owing to the fact that it would be easier to induce diamagnetic current loops that are dominantly along the c-axis of the nanotubes, as compared to those that are in the plane transverse to the c-axis.

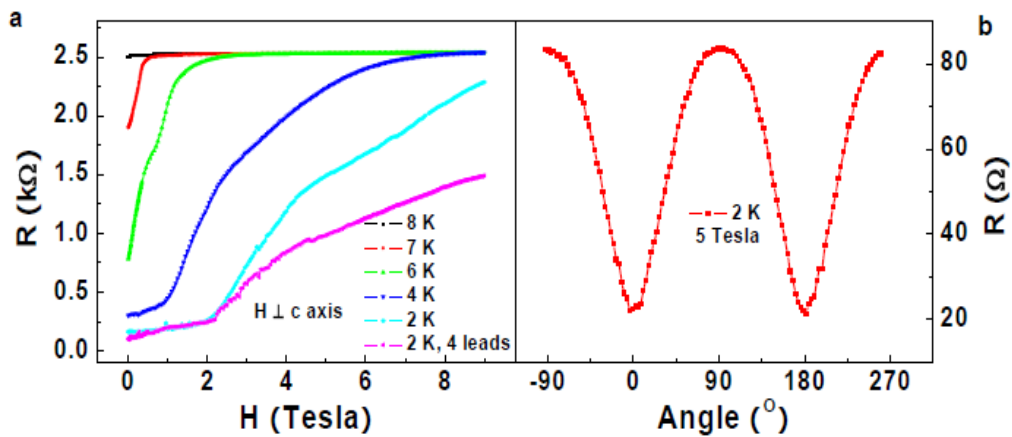

**Figure 3 | Magnetoresistance and its anisotropy.**
**a,** Magnetoresistance (MR) measured at 1µA for field applied perpendicular to the c axis. At 2K, there is a clear transition at ~2.1T for both the two-probe and four-probe data. The two sets of data are about the same magnitude below this field value but diverge above that. This is consistent with the physical picture that global coherence is established at 2K, and a field above 2T would gradually destroy lateral coherence (see Fig. 4b, the green curve).
**b,** MR anisotropy measured on another sample (with 100nA current) that displays qualitatively similar behavior as shown in Fig.2 and Fig. 3a. Here the angle is between the magnetic field and the c axis.

## Physical interpretation

Our nanotube arrays may be considered as composed of 1D superconducting elements which can be either single nanotubes or small bundles of nanotubes each with strong lateral coherence. We characterize each 1D element with a complex Ginzburg-Landau[13] wave function $\psi$ which is a function of the spatial variable $x$ along the c-axis. At temperatures above 7.5K the 1D superconducting condensate can experience strong thermal fluctuations in the form of phase slips, leading to finite resistance. Neighboring elements are coupled via Josephson interaction energy $-J\cos(\varphi_i - \varphi_j)$, where $\varphi_i$ denotes the phase of the wavefunction for the *i*th element (see inset to Fig. 2b), assumed to be constant over a finite segment of length $d$, and $J$ is proportional to the paired (zero temperature) superconducting electron density $|\psi|^2$. We approximate the system as quasi 2D in the lateral plane perpendicular to the c-axis with a plane (film) thickness $d$. In such an approximation the Josephson coupling is equivalent to a 2D spin model with the interaction between neighboring spins expressible as $-J(\vec{s}_i \cdot \vec{s}_j)$, where $\vec{s}$ denotes a unit vector in the lateral plane.

It is well known that there can be vortex excitations in 2D spin systems, consisting of spins that form a closed loop. Vortices of opposite helicities interact as 2D charges and tend to bind together with a logarithmic potential[14,15]. They can be separated at a certain transition temperature, $T_{KT} \approx 2.8J/k_B$, denoted the Kosterlitz-Thouless (KT) transition. As (unpaired) moving vortices can destroy coherence in the lateral plane, its resistance would increase sharply at the KT transition. The appearance of KT transition is noted to be not restricted to thin film superconducting systems; more recently it was also observed in bulk 3D high $T_c$ superconductors that are composed of 2D superconducting atomic layers. These include KT transition observed for vortex excitations within the superconducting layers[16-18], as well as for interlayer vortex excitations involving multiple superconducting layers[19-21].

In our experiments, the surface contact electrodes in the two-probe geometry can effectively detect the lateral resistance change[12], more so than the four-probe geometry. Comparison between the two sets of data shows that above the KT transition the sample's resistance is more anisotropic (see Fig. 2)—whereas the measured two-probe and four-probe resistances are almost the same below 6K (indicating 3D isotropy), they diverge above that.

We attribute the large resistance drop seen at 7.5K (Fig. 2a) to result from a KT transition that establishes quasi long range order in the lateral plane[22], leading to the quench of longitudinal fluctuations and the formation of 3D coherent regions connected by weak links. Global coherence is attained at even lower temperatures.

It is known that the KT transition, which involves only a small entropy change, occurs at a temperature below the main specific heat peak[23-26] that reflects the growth of the transverse coherence. Hence the KT transition attribution is consistent with the previous specific heat results, obtained with a 5T quenching field[3]. In this case, however, there could be synergistic effect between the longitudinal and transverse coherence manifest in the specific heat signal at $T > 7.5K$, that is absent in the standard model of the KT transition. The specific heat signal associated with the KT transition is yet to be observed; it may require better sample quality than that attained in the past, measured with a higher quenching field.

For the transverse-plane KT transition, the effect of the magnetic field is mainly due to its influence on $J$. A perpendicular magnetic field has the effect of suppressing superconductivity as seen in Fig. 3a. In particular, the superconducting electron density and hence the Josephson coupling energy $J$ are reduced. That has the effect of both shifting $T_{KT}$ to lower temperatures as well as diminishing the magnitude of the resistance drop associated with the transition. A parallel field will have a similar effect, although less effective than the perpendicular field as seen in Fig. 3b, owing to the smaller magnetic susceptibility (and hence a larger magnetic penetration length).

## KT transition characteristics

In Fig. 4 we show measured differential resistance plotted as a function of the current. The most striking feature is the low temperature differential resistance gap/quasigap centered at zero bias, which disappears in stages with increasing temperature (4a) or applied magnetic field (4b). A notable aspect of the data is the existence of two resistance plateaus at the large current limit, one at 1kΩ and the other at 2.3kΩ. The latter is associated with the KT transition at 7.5K. A particularly instructive curve in Fig. 4a is the one at 6K (blue), which shows that the bottom of the KT transition's quasigap coincides with the 1kΩ plateau. This is consistent with the physical picture that the 1kΩ resistance is associated with the weak links connecting the 3D superconducting regions. It also justifies treating the 1kΩ plateau as the reference from which the KT transition's lateral plane resistance is to be measured.

In the presence of series resistance internal to the sample, differential resistance measurement can help to clarify the intrinsic nature of the 7.5K transition. As the temperature is lowered below 7.5K, it is seen from Fig. 4a that a triangular-shaped quasigap develops which eventually merges into the 2.3kΩ plateau with increasing current. The resistance in the $I \rightarrow 0$ limit, measured relative to the 1kΩ plateau, is predicted to vary with temperature as[27,28] $R - R_s = 10.8 b R_n \exp\{-2[b(T_{c0} - T_{KT})/(T - T_{KT})]^{1/2}\}$ for $T > T_{KT}$. Here $R_s = 1.06 \text{k}\Omega$ is the lower plateau resistance and $T_{c0} = 7.5K$ is the mean field transition temperature below which there can be nonlinear IV characteristics. In Fig. 2b we show our data to be in excellent agreement with the above prediction. The parameter values obtained are $T_{KT} = 6.17K$, $b = 0.48$, and $R_n = 0.96 \text{ k}\Omega$ for the normal sheet resistance. As the measuring current $I$ increases, we find that at $T < T_{c0} = 7.5K$ there is a distinctive regime where the differential resistance $R$ varies linearly as a function of $I$, implying $V \propto I^2$. The progressive variation from a constant $R$ ($V \propto I$) at 7.5K to a triangular quasigap ($V \propto I^2$) and then to a more rounded gap at below 6.5K, is consistent with the KT transition behavior of $V \propto I^\alpha$ with an $\alpha$ varying from 1 (at the mean field transition temperature $T_{c0}$) to 3 (at $T_{KT}$) or larger with decreasing temperature[23,29]. From our differential resistance data $\alpha = 3$ occurs at $6K < T < 6.5K$, which agrees well with our earlier estimate of $T_{KT} = 6.17K$.

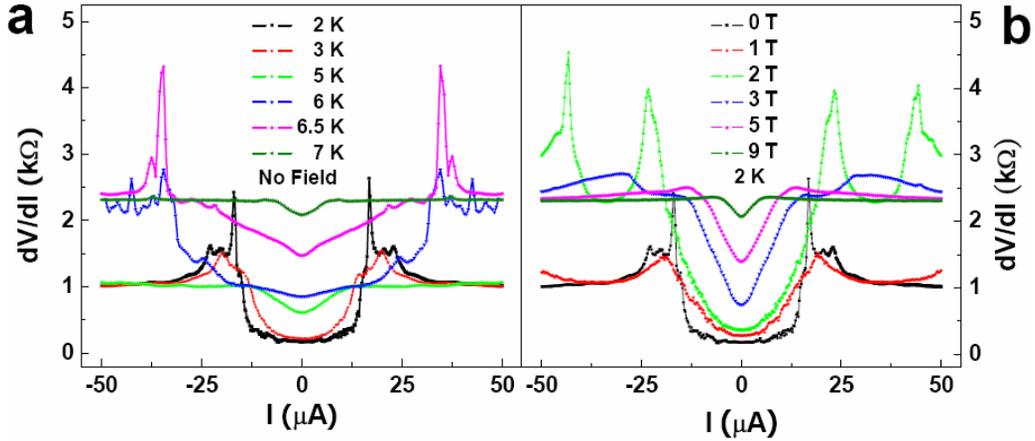

**Figure 4 | Current dependence of the differential resistance.**
**a,** Temperature dependence of the differential resistance plotted as a function of the measuring current. At 2K, the flat platform centered at $I = 0$ is a clear manifestation of the supercurrent gap. It disappears in stages when temperature increases and the global coherence is broken. The 1kΩ plateau is associated with weak links that connect the 3D superconducting regions. Above 6K, lateral coherence deteriorates, and the shape of the quasigap is a reflection of the KT transition's nonlinear *I-V* behavior at $T < T_{c0}$. The triangular-shaped quasigap is noted to be associated with the 2.3 kΩ plateau.
**b,** Differential resistance at 2K under different applied (perpendicular) magnetic fields. The effect of the magnetic field is qualitatively similar to that shown in **a**, but differs in details. At 3T and above the appearance of the triangular quasigap signals the suppression of the KT transition to below 2K.

At 6K and below, a smaller gap is seen to develop which is associated with the 1kΩ plateau. At 2K, a supercurrent gap becomes well-defined, attributable to the establishment of global coherence. The sharp peaks at the boundaries of the supercurrent gap arise from the existence of a critical current density at which voltage first appears (when the weak links are overcome). This general behavior is seen to be repeated in Fig. 4b, in which the applied magnetic field may be viewed as having the effect of decreasing the transition temperature. In particular, the switching of the plateau resistance at 2T, from 1kΩ at below 2T to 2.3kΩ at 2T, is noted to be also reflected in the behavior shown in Fig. 3a, where the curve associated with 2K has a kink at ~2T, beyond which the MR increases with a much larger slope. From the appearance of the triangular quasigap at magnetic fields above 2T, we conclude that a 3T field is sufficient to suppress $T_{KT}$ to below 2K, but $T_{c0}$ can remain above 2K even at 9T. An overall phase diagram is presented in the Supplementary section. It should be noted that the peaks seen in Fig. 4b for the 2T curve (in light green) remain unexplained. Further studies are required to clarify their origin(s).

**Concluding remarks**
We have shown that by measuring 4-Angstrom carbon nanotubes embedded in AFI zeolite at *close electrode separations* the superconducting behavior of coupled nanotube arrays can be clearly delineated through resistance versus temperature, resistance versus magnetic field, and I-V characteristics. The necessity of using close electrode separation, and the related consequence of small room temperature resistance of the samples (in which the superconducting behavior is manifest), indicate that impurities and defects present in most conducting nanotube samples may be responsible for masking their intrinsic characteristics at the scale of 0.5μm or larger. This is reasonable as impurities and defects are much more effective in localizing electrons in 1D than they are in higher dimensions.

While superconductivity in carbon nanotubes is now beyond reasonable doubt, results presented in this work also open up some intriguing questions requiring further investigations, among them the fundamental issue of the pairing mechanism in carbon nanotubes. Research on these problems represents theoretical and experimental challenges to be further pursued.

**Methods**

The 4 Angstrom carbon nanotube-zeolite composites were prepared by first heating the crystals in 0.3 atmosphere of oxygen and 0.7 atmosphere of $N_2$ or Ar at $580^o$ C for 4 hours. The purpose of this initial heating stage is to remove the precursor—tripropylamine—that was present in the pores of the as-made crystals. At the end of the first heating stage the crystals were transparent with no Raman signals for carbon-carbon bonds. Subsequently the 0.3 atmosphere of $O_2$ was replaced by ethylene and the crystals were heated at the same temperature for the same duration. The resulting crystals show strong optical polarization anisotropy with Raman radial breathing modes at both 510 cm$^{-1}$ (for the (4,2)) and 550 cm$^{-1}$ (for the (5,0)). In particular, the peak at the 550 cm$^{-1}$ is about 10% the height of the G-band at 1600 cm$^{-1}$ for the C-C bonds[2].

The transport measurements were carried out in the Quantum Design PPMS, with a 2.1$\Omega$ series resistance. Both resistance and differential resistance were measured using Keithley 2182A Nanovoltmeter and 6221 AC/DC current source.

**References and footnotes**

**Acknowledgements**

We thank Bei Zhang, Fuyi Jiang and Milky Tang for technical support. P.S. thanks Qiucen Zhang for helpful discussions. This work was supported by the Research Grants Council of Hong Kong Grants HKUST9/CRF/08, CA04/04.SC02, plus DSC104/05.SC01 and VPAAO04/05.SC01.

**Supplementary information**

*Magnetic field—temperature phase diagram of coupled superconducting nanotube arrays*

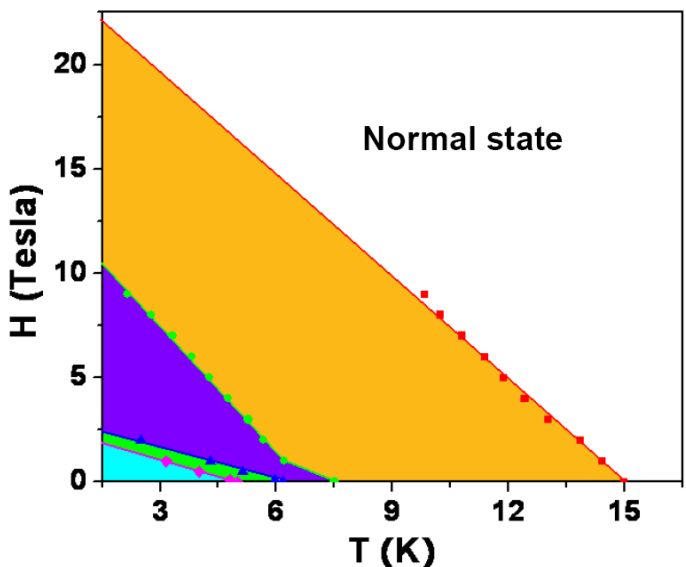

In the phase diagram above, yellow denotes the 1D fluctuating superconductor regime, the green line denotes $T_{c0}$ and the blue line denotes $T_{KT}$ associated with the KT transition. Area colored by violet is the regime where one expects to see nonlinear IV characteristics, owing to vortex excitations. Green is the regime in which the sample is characterized by inhomogeneous 3D superconducting regions connected by (normal) weak links. The bottom left corner is the regime of global coherence. Here symbols are data; straight line extrapolations are used to delineate the different regimes.